\documentstyle[prbbib,prb,twocolumn,aps,epsfig]{revtex}
\begin{document}

\title{ Exact results and mean field approximation
for a model of 
molecular aggregation}

\author{Daniel Duque$^1$ and Pedro Tarazona$^{1,2}$}

\address{
$^1$Departamento de F\'{\i}sica Te\'orica de la Materia
Condensada (C-V) and
$^2$Instituto Nicol\'as Cabrera, \\
Universidad Aut\'onoma de Madrid, E-28049 Madrid, Spain}

\date{\today}
\maketitle

\begin{abstract}

We present a simple one-dimensional model with molecular 
interactions favouring the formation of clusters with a
defined optimal size. Increasing the density, at low 
temperature, the system goes from a nearly ideal gas 
of independent molecules to a system with most of the molecules 
in optimal clusters, in a way that resembles the formation
of micelles in a dilution of amphiphilic molecules, at the
critical micellar concentration. Our model is simple enough
to have an exact solution, but it contains some basic features
of more realistic descriptions of amphiphilic systems:
molecular excluded volume and molecular attractions which
are saturated at the optimal cluster. The comparison between
the exact results and the mean field density functional 
approximation suggests new approaches to study the more
complex and realistic models of micelle formation;
in particular it addresses the long-standing controversy
surrounding the separation of internal degrees of freedom in the
formulation of cluster association phenomena.

\end{abstract}

\pacs{ }

\section{Introduction}

 The study of systems with strong effects of molecular 
aggregation sets problems of fundamental and applied 
interest.  One of the most important examples are the
dilute solutions of amphiphiles in water
\cite{Gener1,Gener2,Gener3}. 
The polar head in the molecule
attracts the water, which is repelled by the
hydrocarbon tails in the same molecule, and
the frustration of these opposite
tendencies leads to the formation of micelles, droplets with
hundreds or thousands of 
molecules, which have their polar 
heads on the surface
(in contact with the water) and the interior filled with the
hydrocarbon tails. The radius of the micelles is limited by 
the length of the molecular chains, so that the droplets
cannot grow far beyond it as spherical drops. 
From a theoretical point of view, micellar 
solutions could be regarded as extreme cases of non-ideal 
solutions, with strong molecular correlation described as
microscopic clusters (micelles). However,
the large number of molecules in each 
micelle makes impracticable the use of the virial expansion
as with ordinary dilutions \cite{hill}.  The alternative is to 
split the description of the systems into two levels: first the
description
within some approximate scheme, of a single cluster of $k$ 
molecules to get its {\it internal} grand potential energy, $\Omega_k$,
and then the whole system is regarded 
as an ideal mixture, with the density of clusters proportional to
the {\it internal} Boltzmann factor 
\begin{equation}
\rho_k \sim \exp[-\beta \Omega_k],
\label{sim}
\end{equation}
with $\beta=(k_b T)^{-1}$. 
There remains controversy about the 
proportionality factor, which should provide the units of 
inverse volume to $\rho_k$ in Eq. (\ref{sim}),
and how this factor may depend on the cluster size and 
mass \cite{Morse,McMullen};
see also the series of articles by Reiss {\it et al.} \cite{Reiss},
and references therein.
The indeterminacy in this factor is directly associated to
the problem of 
how to define the {\it internal} grand-potential energy \cite{ours};
when the
process of micelle formation is described as a chemical
reaction the problem is transferred to the usual indeterminacy
in the {\it standard} chemical potential  \cite{hill,Ben}.
The problem may be avoided assuming that the dependence
of $\rho_k$ with the temperature and the total concentration
of amphiphiles is dominated by the exponential factor.  For
$\Omega_k >0$ micelles of size $k$ are scarce, while for
$\Omega_k<0$ they should be abundant; the {\it critical micellar
concentration} (CMC) is then associated with conditions at which
the minimum value of $\Omega_k$, for the optimal size, $n$, becomes 
zero. This semi-quantitative analysis provides useful 
information from empirical thermodynamic approximations for 
$\Omega_k$, with surface and volume contributions, which 
may include the qualitative effects of molecular shape
\cite{Gener3} and the role of the flexible chains \cite{Gener2,Ben}.
However, attempts to develop a quantitative link between 
approximated (mean field) descriptions of an aggregate at
microscopic level, and the density $\rho_k$ of these aggregates
at equilibrium \cite{Morse,ours}, requires consistent microscopic 
approximations for the prefactor in Eq. (\ref{sim}).
In this paper we explore the
problem with a simple model for molecular aggregation that may
be solved both exactly and with a mean field density functional
approximation (MFA) comparable to that used in more realistic 
models \cite{ours}. The comparison allows us to establish
a workable
approach linking the two levels of statistical description in
these systems, to be used in more sophisticated models which do
not have exact solution.

\section{ Simple model for molecular aggregation}

\begin{figure}
\begin{center}
\epsfig{file=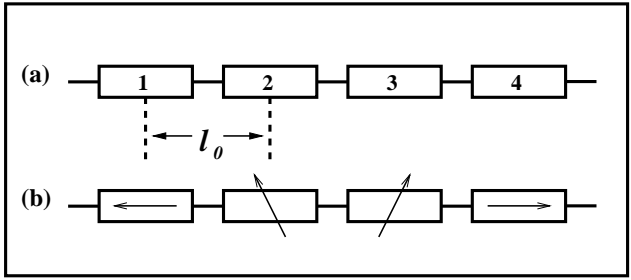,width=8.5cm}
\label{f1}
\caption[]{
Sketch of an optimal molecular aggregate in our
one dimension model with $n=4$. The hard rods are separated by
the optimal distance, $\l_0$, and the internal variable $\xi_i$
takes consecutive values from $\xi_1=1$ to $\xi_n=n$ as indicated
by the numbers in each rod in (a). In the lower
representation (b), the internal variable is represented by
an angular variable with half-clock $n$ possible values, to
give a more intuitive idea of how the non-cyclic character of
$\xi_i$ prevents the grow of aggregates with more than $n$ rods.
}
\end{center}
\end{figure}

 Our model for molecular aggregates is a system of hard rods,
with length $\sigma$, moving along the $X$ axis. Consecutive
rods have positions $x_{i+1}\geq x_i+\sigma$, and each rod has a
discrete internal degree of freedom $\xi=1,...,n$.
Besides the hard-core repulsion there is an interaction energy
between nearest neighbour  pairs,
$\phi(x_{i+1}-x_i,\xi_{i+1}-\xi_i)$
restricted to the range $\sigma \leq x_{i+1}-x_i \leq 2 \sigma$. 
As a simple model with this behaviour we take
\begin{equation}
 \beta \phi(x,\xi)= K \ \xi \  {{ (x - \sigma)(x - 2 \sigma) } 
\over {\sigma^2}}
\label{phi}
\end{equation}
if $\xi \equiv \xi_{i+1}-\xi_i= \pm 1$ and zero otherwise.
The dimensionless parameter $K$ 
gives the inverse temperature, in units of the interaction 
strength.  The optimal aggregate, represented in Fig. \ref{f1}, 
is made of $n$ rods with    
consecutive values of $\xi_i$ from $\xi_1=1$ to $\xi_n=n$,
keeping relative distances between neighbours close to
$\l_0=3 \sigma /2$, to be at the minimum of $\phi(x,\xi)$.
The non-cyclic character of the internal
variable $\xi$ precludes the growth of larger aggregates, 
playing (in a rather different way) the same role as the
chain-packing constrains in real amphiphiles.
Small fluctuations around the optimal distance $x_{i+1}-x_i=l_0$
still have a cohesive energy and provide some internal 
entropy for the aggregate. The aggregates compete with
fully disordered configurations, which maximize entropy but
do not get cohesive energy from Eq. (\ref{phi}).

\subsection{Exact results}

\begin{figure}
\begin{center}
\epsfig{file=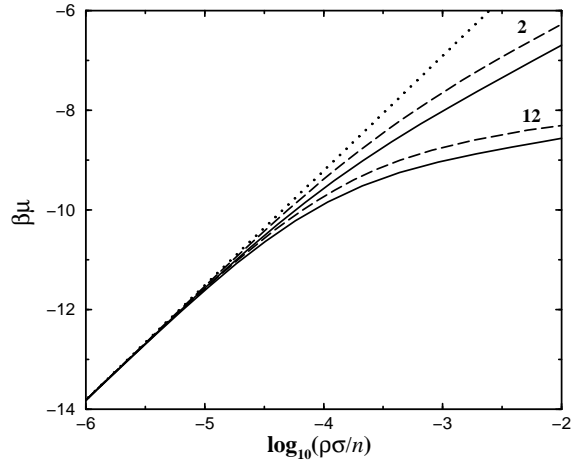,width=8.5cm}
\label{f2}
\caption[]{
Isothermal representation of the chemical potential
versus the molecular density in our model for $K=20$. The full
lines are the exact results for $n=2$ and $n=12$. Broken lines
are the results with the mean field approximation. The dotted
line is the ideal gas limit.
}
\end{center}
\end{figure}

 The exact properties of the model are easily obtained \cite{Henderson}
in a statistical ensemble with constant temperature, $T$, and
pressure,
$p$, in terms of a $(n \times n)$ transfer matrix with elements,
$ {\cal M}_{j,k}(x) = \exp[ -\beta (\phi(x,k-j) - p x)] $
for $j,k=1,..,n$. The integral of ${\cal M}_{j,k}(x)$ from $x=1$
to
infinity gives the elements of the transfer matrix for the
internal
variable, ${\cal Q}_{j,k}$, and its largest eigenvalue
$\lambda_{max}$ 
provides the thermodynamic properties of the system, through the
chemical
potential $ \beta \mu(T,p)= - \ln(\lambda_{max})$, and 
the density of rods, $\rho(T,p)=(\partial p / \partial \mu)_T$.
The isotherms $K=20$ for $n=2$ and $n=12$ presented in Fig. \ref{f2},
give $\beta \mu$ versus $\rho\sigma/n$ (with logarithmic scale). Very
diluted systems present ideal gas behaviour, with curves of slope
$1$  in the figure.
When the molecular association becomes important 
the curves change to have slope $1/n$, which corresponds to
an ideal mixture of $n$-molecule clusters, until the packing
effects show up and the slopes increase.

Any property measuring the correlation structure can also be
obtained
from ${\cal M}(x)$ and ${\cal Q}$. In particular, the density of
aggregates, which is a property of $n$-rod correlations, is 
\begin{equation}
  \rho_n= \rho  \lim_{ N \rightarrow \infty} 
{{ \mathrm{Tr}[{\cal Q}^{N-n+1} {\cal W}]} \over {\mathrm{Tr}[{\cal Q}^N]}}, 
\label{rhoe}
\end{equation}
where the matrix ${\cal W}$ depends on our exact definition for
the aggregate. In the following we define the {\it full aggregate}
(or {\it micelle} in a rather loose sense of the word)
as any group of $n$ rods with $\sigma \leq x_{i+1}-x_i\leq 2 \sigma$, and 
$\xi_{i+1}-\xi_i=1$, starting with $\xi=1$, so that every bond
gets some cohesion energy from Eq. (\ref{phi}). With this definition 
we have
\begin{equation}
{\cal W}_{j,k}= \delta_{j,1} \delta_{k,n} 
\prod_{i=1}^{n-1} \int_{\sigma}^{2 \sigma} dx  {\cal M}_{i,i+1}(x).
\label{W}
\end{equation}
Slightly different definitions of {\it micelle} would lead to other
forms for
${\cal W}$, but they would change $\rho_n$ very little, as long as
they include the optimal and nearby configurations.
 
\begin{figure}
\begin{center}
\epsfig{file=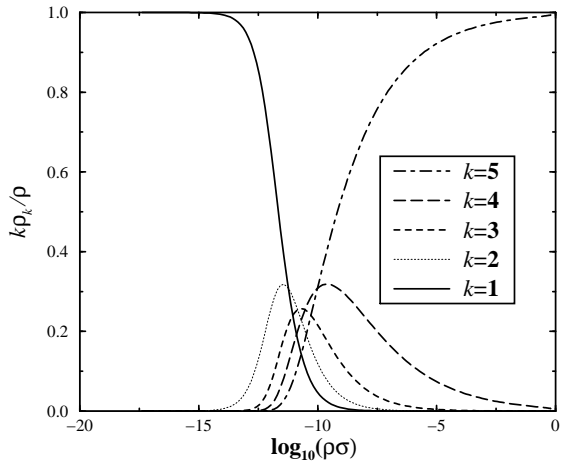,width=8.5cm}
\label{f3}
\caption[]{
Proportion of rods forming part of aggregates of
different size in our model for $n=5$, as a function of the
molecular density along the isotherm $K=60$. The curve with $k=1$
corresponds to isolated molecules, the curve with $k=n$ gives
the proportion of rods in full aggregates.  The {\it critical
concentration} region is  located around the point where
$\rho_1=n \ \rho_n$, and has high relative concentration of
incomplete aggregates, with $2 \le k \le n-1$.
}
\end{center}
\end{figure}

 In a similar way we may calculate the density, $\rho_k$, of
{\it incomplete aggregates}, with  $k < n$ rods, and the density $\rho_1$ 
of isolated rods, which do not get any cohesion from Eq. (\ref{phi}).
In Fig. \ref{f3} we present $k \ \rho_k/ \rho$, i.e., the 
proportion of rods forming part of cohesive clusters of
different sizes, as a function
of the total density $\rho$, along an isotherm. In qualitative 
agreement with real systems of amphiphilic molecules, 
there is a low concentration regime, with very few micelles,
where $\rho \simeq \rho_1$, and a
higher concentration regime, dominated by clusters of the
preferential size $n$, so that $\rho \simeq n \ \rho_n$.  
The density at the transition between the 
two regimes represents the {\it critical micellar concentration}
(CMC, again in a loose sense of the term), which
may be defined at the point where $\rho_1= n \ \rho_n$, so that
there are as many isolated molecules as molecules forming part
of complete micelles.
In the neighbourhood of the CMC there are significative amounts of 
incomplete aggregates, with size between $2$ and $n-1$. The CMC
corresponds to the region where the slopes of the curves in Fig. \ref{f2}
change from $1$ to $1/n$. These changes   
are smoother in our simple, one-dimensional model than
in real amphiphilic systems, but as the CMC is never a true 
phase transition, the use of a one-dimensional model does 
not change the quantitative behaviour of the system. 

\begin{figure}
\begin{center}
\epsfig{file=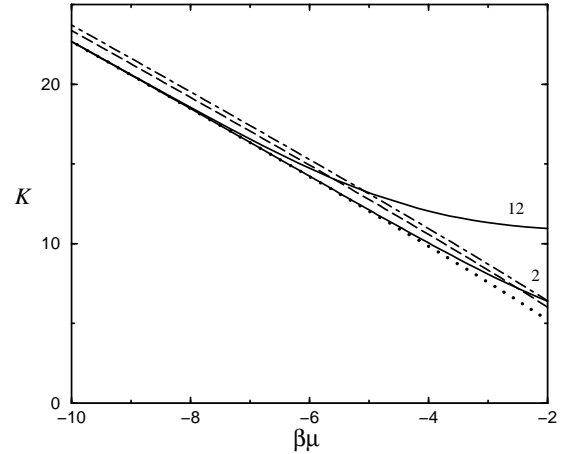,width=8.5cm}
\label{f4}
\caption[]{
The {\it critical concentration} in our model represented
in terms of the parameter $K$ ( or adimensional inverse temperature)
versus the chemical potential $\beta \mu$. The full lines are the exact
results for $n=2$ and $n=12$. They share the same asymptotic
limit at low temperature ($K \gg 1$), as given by Eq. (\ref{mucmc})
and represented by the dotted line. The results with the mean field
approximation are given by the dashed (n=12)  and the dot-dashed
(n=2) lines. They have an asymptotic shift respect to the
exact results which decreases with the aggregate size $n$.
}
\end{center}
\end{figure}

 At low temperatures ($K \gg 1$) the asymptotic analysis of our 
exact results gives the CMC at
\begin{equation}
\beta \mu_{c}= - {1 \over 2}  \left [ K + \ln \left( {\pi \over{
2 K}} \right) \right],
\label{mucmc}
\end{equation}
independent of the optimal cluster size $n$.
In Fig. \ref{f4} we compare Eq. (\ref{mucmc}) with the
exact results obtained from the numerical solution of Eq.  (\ref{rhoe},
\ref{W}). The deviations from the asymptotic limit
at low $K$ (i.e., high temperature)
represent the effects of the molecular packing, which    
force the relative distances between neighbour 
rods below the optimal value $l_0$, with a
compromise between the tendency to build 
the optimal aggregates 
and the packing constrains (which push the rods towards the
distance $\sigma$ given by the hard core).
In real systems of amphiphilic molecules, this compromise
leads to very complex phase diagrams, with
lamellar, hexagonal, cubic phases \cite{Gener1,Gener3}. In our
one-dimensional system with short range interactions, however, the exact 
result cannot include any phase transitions.

\subsection{Mean field approximation}

\begin{figure}
\begin{center}
\epsfig{file=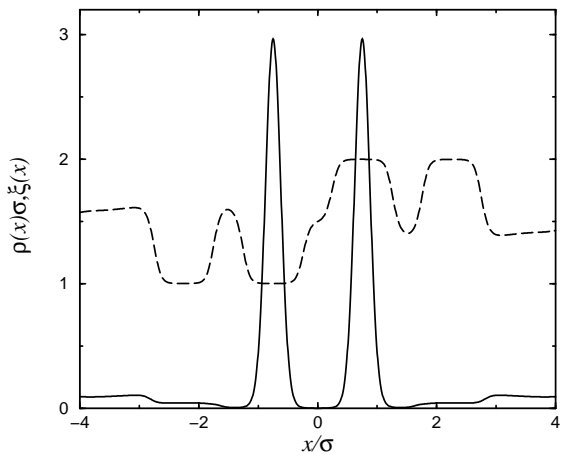,width=8.5cm}
\label{f5}
\caption[]{
The mean field representation of an aggregate as
an inhomogeneous density distribution which is a local
minimum of the grand potential energy in our model with $n=2$, for
$K=16$ and bulk density $\rho_1=0.02 \sigma^{-1}$.
The full line is the total density with two narrow peaks,
and the dashed line is the internal order parameter $\xi(x)$ which is
nearly saturated at the value $\xi=1$ on the left peak and $\xi=2$
on the right one.
}
\end{center}
\end{figure}

We now consider the same model with a density
functional approximation, to describe the aggregates as
mesoscopic self-maintained inhomogeneities in a homogeneous 
bulk phase,
corresponding to local minima of the grand-potential energy
as a functional of the distribution of rods with each value of
the internal parameter, $\rho(x,\xi)$,
\begin{eqnarray*}
\Omega[\rho(x,\xi)]= {\cal F}_{hr}[\rho(x)] - \mu \int dx \rho(x)
+   
\end{eqnarray*}
\begin{eqnarray}
 \sum_{j,k=1}^{n} \int \int 
_{x \leq x'} dx dx'  \phi(x'-x,k-j) \rho(x,j) \rho(x',k),
\label{Omfa}
\end{eqnarray}
were ${\cal F}_{hr}$ is the exact free energy density functional
for 
hard rods \cite{hrods}, which depends only on $\rho(x)=\sum_\xi
\rho(x,\xi)$, and the second line is the mean field (mf)
approximation for
the interaction Eq. (\ref{phi}). This treatment of the model is
equivalent to
what can be done for more realistic models in three dimensions
\cite{ours}. As in those cases, $\Omega[\rho(x,\xi)]$ in Eq.
(\ref{rhoe})
always has a local minimum, with value $\Omega_o$,
for a homogeneous density distribution
$\rho(x,\xi)=\rho/n$, in which, for our model, the contribution
of 
$\phi(x,\xi)$ vanishes, and the system follows the thermodynamics
of 
a hard rod system. However, at low temperature other local minima
may appear, with value $\Omega_o +\Omega_k$, in which the
density distribution $\rho(x,\xi)$ is
made of a series of $k$ narrow peaks, over the uniform background
density (Fig. \ref{f5}). The distance between the peaks and the 
dependence with $\xi$ 
corresponds to configurations close to the optimal aggregate.
This mean field representation
is equivalent to what is found in more realistic
models of micelles in three dimensions. As in those cases, 
the value of $\Omega[\rho(x,\xi)]$ is degenerate with 
respect to a rigid translation of the micelle, which
may be located at any point over the uniform background. The
density of this background represents within the mean field
approximation the density of monomers, $\rho_1$, 
and it may be used as a way to 
control the chemical potential, $\mu=\mu_{hr}(\rho_1,T)$. This 
mean field approximation neglects the entropy associated with the
position of the micelle, and the second step of the statistical
analysis corresponds precisely to include it to obtain the
aggregate density,
$\rho_k$, from their grand potential excess $\Omega_k$,
like in Eq. (\ref{sim}). Our simple model, with exact results for
$\rho_k$, allows us to explore workable approximations for the
proportionality factor in Eq. (\ref{sim}). 

The degeneracy of $\Omega[\rho(x,\xi)]$ with respect to the
position of
the micelle also implies the existence of local minima with two
or more micelles well separated along the system, and with an
excess grand potential which is $\Omega_k$ times the number of
micelles.
Some weak effective attraction may appear between
micelles, mediated by the monomer density, $\rho_1$, and as
the chemical potential increases, the absolute minimum of 
$\Omega[\rho(x,k)]$ goes to dense periodic structures.

\section{Ideal mixtures of molecular aggregates}

\subsection{General approach}

  In this section we consider the generic problem of how to
obtain
the global behaviour of a system from a mean field description
of 
the molecular aggregates in the limit of low total density,
$\rho$. We use the notation appropriate for 
systems in three dimensions to get general expressions, which
we later apply to our simple model in one dimension. 
We assume that there may be several types of molecular
aggregates, and we label them by their number of molecules $k$.
There is always some arbitrariness, at this level of description,
in what
is called an {\it $n$-molecule aggregate}, but the existence of
micelles
with a well defined size and structure makes it easy to find
{\it natural} choices, like the one used in our simple model leading to
(2--3). As discussed above, the use of reasonable variants would 
have very little effect in the results. 
 
In  the limit of low total density 
the system may be considered as an ideal gas of aggregates. The
partition function of an ideal mixture with $N_k$
aggregates of size $k$, 
and a total number of molecules $N=\sum_k k N_k$,
is factorized into the contributions of each aggregate,
\begin{equation}
Z(T,V,\{N_k\})= \prod_k {{ [Z_k(T,V)]^{N_k} } \over {N_k!}},
\label{Z}
\end{equation}
where $Z_k(T,V)$ is the canonical partition function of the 
$k$-molecules aggregate, as a function of the
temperature and the total volume of the system, $V$.
This partition function may be factorized into a translational
part,
proportional to the volume, and the internal part, $Z_k^{int}(T)$
which is independent of $V$. We make it explicit as 
\begin{equation}
Z_k(T,V)= e^{- \beta E_k} \ {{ V  \ (v_k)^{k-1}} \over 
{\Gamma^k } } = {V \over \Gamma} \ Z_k^{int}(T), 
\label{Zn}
\end{equation}
where we have separated the Boltzmann factor, with the mean
energy
of the aggregate $E_k$, and the phase space volume with a factor
$V$. The $k-1$ factors $v_k$ (for which Eq. (\ref{Zn}) may be
considered a
definition) include the phase space volume associated with the
relative positions of all the molecules in the aggregate and any
rotational or internal degree of freedom of the molecules, as
well as any combinatorial factor.  The
total configuration integral over positions has dimensions of
volume to the power $k$, which are cancelled by the factor $\Gamma^k$
in the denominator, where $\Gamma$ is the {\it unit cell volume}
in the configuration phase space. The matching between classical
and quantum statistics may be used to fix $\Gamma$, e.g., for a 
monatomic ideal gas with atomic mass $m$, 
the integral over the momenta  
give $\Gamma=\Lambda^3$, in terms of the {\it thermal wavelength}, 
$\Lambda= \hbar ( 2 \pi \beta / m)^{1/2}$.
However, within the range of validity for the classical
statistical
mechanics all the results related to the positions
of the molecules, like $\rho_k$ or any molecular correlation 
structure, will be invariant if $\Gamma$ 
is multiplied by an arbitrary factor (or if the molecular mass 
$m$ is changed).
Thus, we keep the term $\Gamma^k$ in Eq. (\ref{Zn}), to get a
dimensionless
partition function $Z_k$, but we know that $\Gamma$ and $m$  
have to disappear in any final expression for the 
density of aggregates. By taking $\Gamma$ as an arbitrary constant
(independent of $T$) we neglect the molecular kinetic energy,
which would add (for each classical degree of freedom)
a contribution $(2 \beta)^{-1}$ to the internal energy, without changing
any relevant result here. 

  The mean number of $k$-molecule aggregates in the system,
$N_k$, 
is easily obtained from Eq. (\ref{Z}) in the grand-canonical
ensemble,
$N_k=- \partial (Z_k \ \exp(k \beta \mu ) ) / \partial \beta E_k$,
and the factor $V$ in Eq. (\ref{Zn}) provides a well defined density
of clusters
\begin{equation}
\rho_k \equiv {{ N_k} \over V}={1 \over v_k} \exp\left[ - \beta
(E_k- k \mu) - 
 k \ln(\Gamma/v_k)  \right].
\label{rhon}
\end{equation}
The total density is then given by
\begin{equation}
\rho(T,\mu) \equiv { N \over V} = \rho_1 + \sum_{k>1} \ k \
\rho_k,
\label{rhot}
\end{equation}
where we have separated the density $\rho_1$ of isolated
molecules,
which are not considered to form part of any aggregate. For a
given total density, $\rho$, the chemical potential may be written as
\begin{equation}
 \mu=   \mu_{id} + 
 \Delta \mu= \beta^{-1} \ln(\rho \Gamma) +  \Delta \mu(\rho, T), 
\label{muid}
\end{equation}
where we have separated the ideal gas contribution from the
contribution of the molecular interactions. The former takes 
all the dependence on the phase space unit $\Gamma$, through the
form $\beta \mu_{id}= \ln(\rho \Gamma)$. Substitution
of Eq. (\ref{muid}) into Eq. (\ref{rhon}) shows the perfect cancellation of
the $\Gamma$ factors in the density of aggregates, $\rho_k(\rho,T)$ 
when it is written as a function of the total density
and the temperature.

Consider now the description of the same system with a generic 
mean field density functional approximation for the grand
potential
energy $\Omega[\rho]$. There is always a relative minimum of
$\Omega$ for a homogeneous bulk density distribution, which we
associate 
with the density $\rho_1(T,\mu)$ of unaggregated molecules in
Eq. (\ref{rhot}).
There may be other functional local minima, for inhomogeneous 
distribution functions, and each one is associated with a type
of molecular aggregate. Compared with the uniform bulk phase, 
this aggregate has an excess of $k$ molecules (which again we use
to
label the type of aggregate) and an excess Helmholtz free energy,
$F_k^{mf}$. In the limit of a very dilute bulk we may represent
the 
mean field approximation for the partition function on the 
$k$ molecules as 
\begin{equation}
Z_k^{mf}(T) \equiv \exp[-\beta F_k^{mf}] = 
e^{- \beta E_k^{mf}} \ \left( {{ v_k^{mf}} \over 
{\Gamma} } \right)^k
\label{Zmf}
\end{equation}
where $E_k^{mf}$ is the energy given by the mean field
approximation,
and $v_k^{mf}$, defined through Eq. (\ref{Zmf}),
gives the volume in the configuration phase space per molecule
in the same approximation. The qualitative difference between
Eq. (\ref{Zmf}) and the exact result Eq. (\ref{Zn}) is the lack of a
volume factor $V$.
This factor arises from the overall position of the aggregate,
but in our mean field description we are neglecting the
collective
mode associated to $k$-molecules correlations. The translational 
invariance is still reflected
in the degeneracy of the density functional minimum, but it does
not
contribute to $F_k^{mf}$, which is independent of the total
volume 
in the system.

We now make explicit the use of the mean field
approximation for the statistics of the aggregates by using
the exact expression Eq. (\ref{rhon}) for $\rho_k$, with the results
of the
mean field approximation for $E_k$ and $v_k$; the expression in
the
exponential becomes
\begin{equation}
- \beta (E_k^{mf}- k \mu) - k \ln(\Gamma/v_k) =
- \beta F_k^{mf} + k \beta \mu
\label{mfa}
\end{equation}
which is precisely $ -\beta \Omega_k^{mf}$, as in 
empirical treatments, but we now get an explicit approximation 
for the prefactor:
\begin{equation}
\rho_k= {1 \over v_k^{mf}}  e^{-\beta \Omega_k^{mf}}
\label{rhomf}
\end{equation}
where $v_k^{mf}$ may be obtained through Eq. (\ref{Zmf}), 
from the difference between 
the internal energy and the Helmholtz free energy in the same mean
field
approximation,
\begin{equation}
 v_k^{mf}= 
{1 \over {\rho_1}} \  \exp \left [
 \beta  {{E_k^{mf}-\Omega_k^{mf}} \over k} - \beta \Delta \mu(\rho_1) 
\right ]
\label{vmf}
\end{equation}
in terms of the bulk density of unaggregated molecules, the
internal energy $E_k^{mf}$ and the grand potential 
excess $\Omega_k^{mf}$.

 Within this approach we may obtain the thermodynamics and the 
structure of a system with a strong degree of molecular
aggregation
from a mean field description of each separated cluster. For each
local minimum of $\Omega[\rho]$ we get a type of cluster, which
at a given temperature and chemical potential, may coexist with
a bulk of density $\rho_1(T,\mu)$  of uncorrelated molecules.
The total density of the system is given by Eq. (\ref{rhot}), adding
to $\rho_1$ the contribution Eq. (\ref{rhomf}) of each type of aggregate
(i.e., each local minimum of $\Omega$).  The proposed expresion 
Eq. (\ref{vmf}) for the prefactor in Eq. (\ref{rhomf}), is based on
the formal comparison of the exact and the mean field partition 
functions. The factor $V$ of the free translational mode, 
which is missing in the MFA, is added in substitution of an 
{\it average mode} with contribution $v_k^{mf}$.  
There are different {\it recipes} to get $v_k$, and most of them
are based on the phase space unit $\Gamma$ (taken in terms of the
thermal wavelength), which depends on the molecular
mass \cite{Morse,McMullen}, inconsistent with
the use of classical statistics. An empirical recipe  
proposed by us \cite{ours}, was based on the {\it micelle
compressibility} as a way to estimate the contribution of the
average mode.  That approximation was used in the description
of micelles, in a three-dimensional model of amphiphilic systems,
with sensible results. However, it was too sensitive to the 
structure of the aggregates, and it was difficult to
use for {\it crystal micelles} with strong internal structure.

Our present proposal is a consistent thermodynamic evaluation
of $v_k^{mf}$ from the entropy of the aggregate, as given in the
same MFA as used for the grand potential energy $\Omega_k^{mf}$.
This requires getting the
grand potential and the internal energy $E_k^{mf}$
from the same approximation. Obviously,
this level of description is always based on an approximation;
there is not an exact value for the prefactor in Eq. (\ref{sim})
because  the exact description of the system (as we have done
with the simple one-dimensional model) would not give any local
minima of the grand potential energy for inhomogeneous density 
distributions, and there is not an exact value for the exponent
$\Omega_{n}$. In the next subsection we use our simple model of 
molecular aggregation to check the 
precision of this approach by direct comparison with the exact
results, which is obviously not possible in more realistic
models. 

\subsection{ Application to the one-dimensional model}

It is straightforward to use Eqs. (\ref{rhot}, \ref{rhomf} and \ref{vmf}) 
for our simple model in one dimension. The mean field 
approximation Eq. (\ref{Omfa}) has explicit separation between the 
internal energy and the entropic contributions to
$\Omega[\rho]$. At low temperatures ($K \gg1$), we may 
use a Gaussian parametrization of the density distribution:
\begin{equation}
\rho(x,\xi)= {{\rho_1} \over n} +
\sum_{i=1}^m C_i(\xi) \exp[-\alpha_i (x-x_i)^2],
\label{gaus}
\end{equation}
with $C_i(\xi)$, $\alpha_i$ and $x_i$, as free variational
parameters to minimize $\Omega[\rho]$. The homogenous diluted
bulk with density $\rho_1$ corresponds to have $C_i(\xi)=0$
for all $i$ and $\xi$. With the same value of the chemical
potential, $\beta \mu(T,\rho_1)=  \ln(\rho_1 \Gamma/n)$,
we may find other local minima with the structure of
complete or incomplete aggregates, with $k \leq n$ Gaussian
peaks with the correct sequence of the internal 
variable, $C_i(\xi) \simeq ( \alpha_i/ \pi)^{1/2}$, if $\xi=i$ 
and $C_i(\xi) \simeq 0$ otherwise. The distance between  Gaussians
is given by the minimum of the interaction potential Eq. (\ref{phi}), 
$x_{i+1}-x_i=l_0=3 \sigma /2 $, and the Gaussian widths are 
$ \alpha_i \simeq 4 K/\sigma^2$ for the inner Gaussians 
and $\alpha_i \simeq 2 K/\sigma^2$ for those at the ends 
of the aggregates.

 Within this parametrization, with normalized Gaussian peaks,
the excess of internal and Helmholtz free energies of an aggregate
of size $k$ are obtained analytically as
\begin{equation}
\beta E_k^{mf}= -{{k-1} \over 2} K + { k \over 2}
\label{Egaus}
\end{equation}
and
\begin{equation}
\beta F_k^{mf}=- {{(k-1) \ K} \over 2}  +
{k  \over 2}  \ln \left( {{4 K \Gamma^2} \over
{ \pi \sigma^2}} \right)  -\ln(2).
\label{Fgaus}
\end{equation}
We have checked, for the simplest case  $n=2$, that these expressions 
are very close to the results of the full (numerical) functional
minimization of $\Omega[\rho]$, for the relevant region of
$K > 20$ and $\rho_1 < 10^{-2} \sigma^{-1}$.  

From Eqs. (\ref{Egaus} and \ref{Fgaus}), we get  
$\Omega_k^{mf}$ and $v_k^{mf}$ in Eq. (\ref{vmf}), for each of the local
minima
\begin{equation}
\beta \Omega_k^{mf}= - {{(k-1) \ K} \over 2} -
k \ln\left( {{\rho_1 \sigma \pi^{1/2} }\over {2 n K^{1/2}}} 
\right)
- \ln(2)
\label{Ogaus}
\end{equation}
and 
\begin{equation}
v_n^{mf}= 2^{1/n} \left[ {{ e \pi } \over { 4 K}}
\right]^{1/2}  \sigma.
\label{vgmf}
\end{equation}
Both magnitudes
become independent of the phase space unit volume $\Gamma$;
they depend on the parameters of the interaction potential
$n$, $K$ and $\sigma$, the latter
providing the only physical length scale in the model.

We now use Eqs. (\ref{Ogaus} and \ref{vgmf}) in Eqs. (\ref{rhot} and \ref{rhomf}) to get 
the MFA result for the total density $\rho$. In Fig. \ref{f2} the results
of this MFA are compared with the exact results along the isotherm
$K=20$ for $n=2$ and $12$. The qualitative trends are similar:
the ideal gas is recovered at very low bulk density, but
there is a smooth transition (around the CMC) 
to a regime in which $\beta \mu$
has a slope of $1/n$, which corresponds to the ideal
gas of  full aggregates. In this regime the MFA goes
nearly parallel to the exact curves, until the packing
effects (neglected at this level of the MFA) become relevant.
The CMC, still defined from $n \rho_n= \rho_1$, is shifted from the
exact asymptotic result Eq. (\ref{mucmc}), by
\begin{equation}
\beta ( \mu_{c}^{mf} - \mu_{c} ) =  
 {{n-2} \over {2 n}} \ln(2) + {1 \over {2 (n-1)}} .
\label{mucmfa}
\end{equation}
In Fig. \ref{f4} we compare the
the exact and the MFA results for the CMC with $n=2$ and $12$. In the 
asymptotic regime the difference is independent of $K$.  
The MFA prediction for the CMC is shifted to larger values 
of the chemical potential: $\beta (\mu_c^{mf}-\mu_c)=1/2$,
for $n=2$, and  $\ln \sqrt{2} =0.346$
for large $n$. Within our approach this shift 
reflects the combination of
the error made by the MFA in the estimation of $\beta E_n^{mf}$ 
(in the asymptotic regime Eq. (\ref{Egaus}) is $1/2$ larger than the 
exact value), and the error associated with the MFA estimation of 
the cluster {\it internal  entropy} required to get $\Omega_n^{mf}$ and 
$v_n^{mf}$. In our simple system we may compare the  latter
with the exact $v_n$, as defined in Eq. (\ref{Zn}). In the 
limit $K \gg 1$ we get 
\begin{equation}
v_n=  \left[ {{ e \pi } \over { 2 K}}
\right]^{1/2} \sigma
\label{vgaus}
\end{equation}
 This exact result is independent of the optimal cluster size, 
$n$, and it is proportional to $T^{1/2}$ (through the parameter
$K \sim T^{-1}$). The value given by the mean field approximation
Eq. (\ref{Zmf}) shows the same dependence with the temperature 
and becomes identical to the exact value for $n=2$. For
larger $n$ the result Eq. (\ref{vgaus}) goes like $2^{1/n}$
and saturates at large $n$ with a discrepancy of 
$\sqrt2$ with the exact result. As comparison, the identification
of $v_n$ with the thermal wavelength of the aggregate would
lead to   a qualitatively different dependence on $n$ and $T$,
$v_n= \Lambda_n= \hbar ( 2 \pi \beta / ( n \ m))^{1/2}
\sim  (n \ T)^{-1/2}$, and with an unphysical dependence on 
the molecular mass and $\hbar$.
In our simple, one-dimensional, model the aggregates have a
strong crystalline structure, described by the narrow Gaussian 
peaks in the MFA density distribution, which becomes very rigid
at $K \gg1$. This character of the aggregates is 
responsible for the complete failure of the empirical recipe
for $v_n$ which was proposed by one of us \cite{ours} in terms
of an {\it aggregate compressibility}. That empirical recipe was
used in a three-dimensional model, with much softer aggregates
in which every molecule could move over the whole aggregate,
like in a liquid drop. The application to our present model
leads to a problematic formulation of $v_n$ (as the square root of
a negative number) and should be fully discarded.

\section{Conclusions}

  We have approached the study of ideal mixtures of molecular 
aggregates from a mean field approximation for the structure
and internal free energy of the isolated aggregates. The 
general analysis in Section III A suggests a consistent approximation
to get the missing factor in the relation Eq. (\ref{sim}) between 
the internal grand-potential energy of a cluster and its 
abundance in the equilibrium solution. The simple
one-dimensional model of molecular aggregation presented
in Section II, provides a test for our approach. As in more realistic
models, it includes hard-core repulsions and 
saturating attractive interactions, which set an optimal cluster
size $n$. The exact solution of the model shows qualitative 
agreement with the behaviour of a system of amphiphilic molecules
around the critical micellar concentration. Along an isotherm the
system goes from an ideal gas of independent molecules (at very
low chemical potential) to an ideal gas of optimal clusters or
micelles, at higher chemical potential. This transformation
takes place over a range of $\mu$ which becomes narrower as 
$n$ increases (although at a very slow rate because of the
one-dimensional character of the model). The same model is
then solved within a mean field approximation equivalent
to what may be done in tridimensional models \cite{ours}, and
the consistent mean field  result for the density of
clusters Eqs. (\ref{rhomf} and \ref{vmf}) may be compared with
the exact result.  The comparison serves to validate  
our proposed approximation Eq. (\ref{vmf}) for the prefactor $v_n$ in
the density of aggregates. As it should be expected, this prefactor
is not related to the thermal wavelength of the aggregate,
which would imply an unphysical dependence on the molecular 
mass and the wrong dependence on the temperature and the size of the 
aggregate. Our approximation reproduces the exact dependence on the 
temperature, and for large aggregates $n \gg 1$ it recovers
(asymptotically) the exact independence on $n$.  The numerical
discrepancy between the exact and the MFA values of $v_n$ should
be regarded as inherent to the mean field description 
of the aggregates. The thermodynamic consistency of the approximations
for the exponent $\Omega_n$ and the prefactor $v_n$ in the density of 
aggregates is the new feature of our proposal, and it could 
be used for any, more  realistic, model of mesoscopic aggregates.

\section{Acknowledgements}

We are grateful to Enrique Chac\'on and Andr\'es Somoza for
their useful comments and accurate criticism.
  
This work has been supported by the Direcci\'on General de
Investigaci\'on 
Cient\'{\i}fica y T\'ecnica of Spain, under Grant No. PB94-005-C02,
and by the Comunidad Aut\'onoma de Madrid, under Grant No. FPI-1995.

\end{document}